\documentstyle[12pt]{article}
\newcommand{\be}{\begin{eqnarray}}
\newcommand{\ee}{\end{eqnarray}}

\title{Nucleon-Nucleon Correlations and Six-Quark Cluster Effects in
Semi-Inclusive Deep Inelastic Lepton Scattering off Few-Nucleon Systems}

\author{\\C. Ciofi degli Atti$^{(a)}$ and S. Simula$^{(b)}$\\\\
$^{(a)}$Department of Physics, University of Perugia and\\ Istituto Nazionale
di
Fisica  Nucleare, Sezione di Perugia,\\ Via A. Pascoli, I-06100 Perugia,
Italy\\
$^{(b)}$Istituto Nazionale di Fisica Nucleare, Sezione Sanit\'a,\\ Viale Regina
Elena 299, I-00161 Roma, Italy\\\\}

\date{}

\begin{document}

\maketitle

\abstract{ Semi-inclusive deep inelastic lepton scattering off few-nucleon
systems is investigated assuming that virtual boson absorption occurs on a
hadronic cluster which can be either a two-nucleon correlated pair or a
six-quark bag. In both cases the relevance of nuclear effects on forward and
backward nucleon emissions is illustrated and the differences expected in the
energy distribution of the emitted nucleons are analyzed both at $x<1$ and
$x>1$.}

\newpage

\pagestyle{plain}

{\bf 1. Introduction}

\vspace{0.5cm}

\indent The investigation of inclusive and exclusive deep inelastic scattering
(DIS) of leptons off nuclei can shed light on the origin of nuclear forces and
the short range structure of hadronic matter. Inclusive DIS processes off
nuclei
in the kinematics regions $x<1$ ($x = Q^2 / 2M \nu$ being the Bjorken scaling
variable) have provided a wealth of information on sea and valence quark
distributions in nuclei \cite{DVH}; however, much work remains to be
done in order to unravel the short-range structure of hadronic matter. To this
end, two other types of DIS processes would be of great relevance: i)
inclusive scattering at $x>1$, a process forbidden on a free nucleon, and,
ii) semi-inclusive processes, in which, besides the scattered lepton, another
particle is detected in the final state. As far as the first process is
concerned, present experimental data do not reach sufficiently high values of
the momentum transfer, so that at $x>1$ they are mainly due to quasi-elastic
scattering and not to DIS \cite{DAY}; it is only recently that the deep
inelastic nuclear structure function has been obtained in a narrow range of
$x>1$ ($x<1.3$) \cite{SAV91}. As for the semi-inclusive processes, which could
provide information on the effects of the nuclear medium on quark and gluon
distributions, the relevance of exotic configurations at short nucleon-nucleon
(NN) distances, and the mechanism of hadronization, several experiments have
been performed \cite{MAT89} - \cite{KGA}, which however still lack of a
clear-cut interpretation. In ref. \cite{CS93} semi-inclusive DIS processes
$A(\ell, \ell' N)X$ off complex nuclei have been analyzed within the so-called
spectator mechanism according to which, after lepton interaction with a quark
belonging to a nucleon of a correlated NN pair, the recoiling nucleon is
emitted and detected in coincidence with the scattered lepton. It has been
shown that at $x<1$ the cross section corresponding to such a mechanism
represents a valuable tool to detect the momentum and energy dependencies of
the nucleon structure function in the nuclear medium, whereas at $x>1$ it
provides information on the short range part of light-cone momentum
distributions. In ref. \cite{CAR91} the effects of six-quark ($6q$) cluster
configurations upon nucleon emissions in semi-inclusive DIS processes off the
deuteron have been investigated by considering that, after lepton interaction
with a quark belonging to a $6q$ cluster, nucleons can be formed out of the
penta-quark ($5q$) residuum and emitted forward as well as backward. The aim of
this paper is two-fold: ~ i) to extend to few-nucleon systems the analysis of
semi-inclusive DIS processes off complex nuclei performed in ref. \cite{CS93}
within the spectator mechanism; ~ ii) to  generalize the approach of ref.
\cite{CS93} to the case when the virtual boson is absorbed by a quark belonging
to a colorless $6q$ bag. In case of electron scattering the semi-inclusive
cross section reads as follows
 \be
    {d^4 \sigma \over dx ~ dQ^2 ~ d\vec{p}_2} = \sigma_{Mott} ~ \sum_{i=1}^4
    ~ V_i(x, Q^2) ~ W^A_i(x, Q^2, \vec{p}_2)
 \ee
where $Q^2 = - q^2 = \vec{q}^2 - {\nu}^2 > 0$ is the squared four-momentum
transfer; $V_i$ is a kinematics factor; $W^A_i$ is the semi-inclusive nuclear
response; $\vec{p}_2$ is the momentum of the detected nucleon. Although in this
paper calculations will be presented including all four nuclear responses in
(1), in what follows only the Bjorken limit of the structure function
$F^A_2(x, Q^2, \vec{p}_2) \equiv \nu W^A_2(x, Q^2, \vec{p}_2)$ will formally be
considered, since this suffices to illustrate the relevant features of nucleon
production mechanisms. Within the above-mentioned multiquark cluster picture,
the semi-inclusive nuclear structure function $F^A_2(x, \vec{p}_2)$ can be
written as the incoherent sum of the contributions resulting from virtual
photon
absorption by different multiquark cluster configurations (cf. refs.
\cite{FS81} and \cite{YVHP90}); in particular, when $3q$ and $6q$ clusters are
considered, one has
 \be
    F^A_2(x, \vec{p}_2) = P^{(3q)} ~ F^{A,3q}_2(x, \vec{p}_2) +  P^{(6q)}
    ~ F^{A,6q}_2(x, \vec{p}_2)
 \ee
where $P^{(3q)} (P^{(6q)})$ is the probability to have a colorless $3q$
($6q$) bag inside the hadronic cluster absorbing the virtual photon ($P^{(3q)}
+
P^{(6q)} = 1$). The case $P^{(6q)} = 0$, which corresponds to electron
interaction with a quark belonging to a nucleon of a correlated NN pair
($F^A_2(x, \vec{p}_2) = F^{A,3q}_2(x, \vec{p}_2)$), is considered in Section 2,
where the $3q$ cluster contribution is analyzed within the spectator mechanism.
The opposite case, $P^{(6q)} = 1$, which corresponds to electron interaction
with a quark belonging to a 6q bag  ($F^A_2(x, \vec{p}_2) = F^{A,6q}_2(x,
\vec{p}_2)$), is presented in Section 3, where the differences in the energy
distribution of the emitted nucleons are analyzed both at $x<1$ and $x>1$.
Nuclear effects on backward and forward nucleon emissions have been taken care
of by adopting the extended two-nucleon correlation model \cite{CSFS91},
\cite{CS93}, which takes into account the motion of the center of mass (c.m.)
of the hadronic cluster interacting with the incoming lepton. Finally, the main
conclusions are summarized in Section 4.

\vspace{0.5cm}

{\bf 2. The cross section for the process A(e,e'N)X: spectator mechanism}

\vspace{0.5cm}

\indent Let us consider the process in which a virtual photon interacts with a
nucleon of a correlated NN pair, and the recoiling nucleon is emitted and
detected in coincidence with the scattered electron. Within the impulse
approximation the semi-inclusive nuclear structure function reads as follows
(cf. ref. \cite{CS93})
 \be
    F^{A,3q}_2(x, \vec{p}_2) & = & M \sum_{N_1=n, p} Z_{N_1} \int_x^{{M_A \over
    M} - z_2} dz_1 ~ z_1 ~ F^{N_1}_2({x \over z_1}) ~ \int d \vec{k}_{c.m.}
    ~ dE^{(2)} \\ \nonumber
    & & P_{N_1N_2}(\vec{k}_{c.m.} - \vec{p}_2, \vec{p}_2, E^{(2)}) ~ \delta(M_A
    - M(z_1 + z_2) - M^f_{A-2} z_{A-2})
 \ee
where $Z_{p(n)}$ is the number of protons (neutrons); $\vec{k}_1$ and
$\vec{k}_2$ are initial nucleon momenta in the lab system before interaction
with c.m. momentum $\vec{k}_{c.m.} =\vec{k}_1 + \vec{k}_2$; $\vec{p}_1 =
\vec{k}_1 + \vec{q}$ and $\vec{p}_2 = \vec{k}_2$ are nucleon momenta in the
final state; $F^N_2$ is the structure function of the struck nucleon. In Eq.
(3), $x / z_1$ represents the Bjorken variable of the struck nucleon having
initial light-cone momentum $z_1 = k^+_1 / M$; $z_2 = (\sqrt{M^2 + p_2^2} -
p_2 \cos \theta_2) / M$  is the experimentally measurable light-cone momentum
of the detected nucleon ($\theta_2$ is the detection angle with respect to
$\vec{q}$ and $p_2 \equiv |\vec{p}_2|$); $z_{A-2} = (\sqrt{(M^f_{A-2})^2 +
k_{c.m.}^2} + (k_{c.m.})_{\|}) /  M^f_{A-2}$ is the light-cone momentum of the
residual (A-2)-nucleon system with final mass $M^f _{A-2} = M_{A-2} +
E^*_{A-2}$ and intrinsic excitation energy $E^*_{A-2}$.

\indent The relevant nuclear quantity in (3) is the two-nucleon spectral
function, which represents the joint probability to find in a nucleus two
nucleons with momenta $\vec{k}_1$ and $\vec{k}_2$ and removal energy $E^{(2)}$.
For deuteron it simply reduces to the nucleon momentum distribution and for
$^3He$ to the square of the wave function in momentum space, times the removal
energy delta function $\delta(E^{(2)} - E^{(2)}_{thr})$, with $E^{(2)}_{thr} =
2M + M_{A-2} - M_A$ being the two-nucleon break-up threshold.  In case of
$^4He$ and heavier nuclei, the two-nucleon spectral function is not yet
available in the exact form; however, realistic models taking into account
those features of the two-nucleon spectral function which are relevant in the
study of semi-inclusive DIS processes, have been developed \cite{CS93},
\cite{FS81}. In this work two of these models will be adopted: the first one is
the two-nucleon correlation (2NC) model of ref. \cite{FS81}, based on the
assumption that the c.m. of the correlated pair is at rest, which leads to
 \be
    P_{N_1N_2}(\vec{k}_1, \vec{k}_2, E^{(2)}) = n^{rel}_{N_1N_2}(|\vec{k}_1 -
    \vec{k}_2| / 2) ~ \delta(\vec{k}_1 + \vec{k}_2) ~ \delta (E^{(2)} -
    E^{(2)}_{thr})
 \ee
where $n^{rel}_{N_1N_2}$ is the momentum distribution of the relative motion of
the two nucleons in a correlated pair; the second model is the extended 2NC
model, where the c.m. motion of the correlated pair is properly taken into
account (see ref. \cite{CSFS91}); it yields
 \be
    P_{N_1N_2}(\vec{k}_1, \vec{k}_2, E^{(2)}) = n^{rel}_{N_1N_2}(|\vec{k}_1 -
    \vec{k}_2| / 2) ~ n^{c.m.}_{N_1N_2}(|\vec{k}_1 + \vec{k}_2|) ~ \delta(
    E^{(2)} - E^{(2)}_{thr})
 \ee
where $n^{c.m.}_{N_1N_2}$ represents the momentum distribution of the c.m. of
the correlated pair. It can be seen that in both models the (A-2)-nucleon
system is assumed to be in its ground-state; such an assumption is justified by
the fact that only soft components of the c.m. motion distribution are
considered in (5). As a matter of facts, it can be demonstrated that the
excited states of the residual (A-2)-nucleon system contribute only to the
high momentum tail of $n^{c.m.}_{N_1N_2}$ (cf. refs. \cite{CPS84} and
\cite{MOR91}). It should be pointed out that the extended 2NC model reproduces
the high momentum and high removal energy components of the single-nucleon
spectral function of $^3He$ and nuclear matter, calculated using many-body
approaches, as well as the high momentum part of the single-nucleon momentum
distribution of light and complex nuclei \cite{CSFS91}, \cite{CS}. The relative
and c.m. momentum distributions are normalized according to
 \be
    \sum_{N_1, N_2 = n,p} Z_{N_1} \int d \vec{k}_{rel}
    ~ n^{rel}_{N_1N_2}(|\vec{k}_{rel}|) \int d \vec{k}_{c.m.} ~
    n^{c.m.}_{N_1N_2}(|\vec{k}_{c.m.}|) = S_1 \cdot A
 \ee
where $S_1$ is the probability of finding a nucleon in a correlated NN pair or,
more precisely, the probability that, after the removal of a nucleon from a
nucleus, the residual (A-1)-nucleon system is in any state of its continuum.

\indent Let us first analyze the semi-inclusive process within the 2NC model,
i.e. let us assume that the correlated NN pair is at rest with respect to the
(A-2)-nucleon system ($\vec{k}_{c.m.} = \vec{k}_1 + \vec{k}_2 = 0)$. Inserting
(4) in (3), one obtains the well-known result \cite{FS81} that the
semi-inclusive nuclear structure function is directly proportional to the one
of the struck nucleon, viz.
 \be
    F^{A,3q}_2(x, \vec{p}_2) = \sum_{N_1=n,p} Z_{N_1} ~ n^{rel}_{N_1N_2}(p_2)
    ~ \bar{z}_1 ~ F^{N_1}_2({x \over \bar{z}_1})
 \ee
where
 \be
    \bar{z}_1 = 2 - z_2 - E^{(2)}_{thr} / M
 \ee
is the light-cone momentum of the struck nucleon. Thus, within the 2NC model
(i.e., the correlated NN pair at rest) the semi-inclusive nuclear structure
function factorises into a momentum-dependent nucleon structure function
$F^N_2(x / \bar{z}_1)$ and a nuclear quantity related to the relative momentum
distribution of the correlated pair; more important, the light-cone momentum
$\bar{z}_1$ of the struck nucleon is linked to the measured value of  $z_2$ by
the simple relation given by (8) and, if the free nucleon structure function is
used in (7), nuclear effects reduce to a momentum dependent rescaling of the
argument of $F^N_2$ \cite{FS81}. It should be pointed out that at $x>1$
backward nucleon emission is kinematically forbidden within the 2NC model,
because the condition $\bar{z}_1 \geq x > 1$ implies $z_2 < 1$ only (cf. (8)).

\indent Let us now consider the two-nucleon spectral function given by (5).
By allowing the correlated pair to share its c.m. momentum with the residual
(A-2)-nucleon system, the semi-inclusive nuclear structure function is no
longer
proportional to the nucleon structure function $F^N_2$, but is given by the
following convolution integral
 \be
    F^{A,3q}_2(x, \vec{p}_2) & = & M \sum_{N_1=n, p} Z_{N_1} \int_x^{{M_A \over
    M} - z_2} dz_1 ~ z_1 ~ F^{N_1}_2({x \over z_1}) ~ \int d \vec{k}_{c.m.} ~
    n^{c.m.}_{N_1N_2}(|\vec{k}_{c.m.}|)
    \\ \nonumber
    & & n^{rel}_{N_1N_2} (|{\vec{k}_{c.m.} \over 2} - \vec{p}_2|)
    ~ \delta(M_A - M(z_1 + z_2) - M_{A-2} z_{A-2})
 \ee
It can be seen that not only the motion of the c.m. of the pair destroys
the factorization generated by the 2NC model (cf. (7)), but, moreover, the
light-cone momentum $z_1$ of the struck nucleon depends both upon $z_2$ and
$z_{A-2}$ and, consequently, it cannot be directly related to measurable
quantities.

\indent Within the extended 2NC model nuclear effects on backward and forward
nucleon emissions in the semi-inclusive DIS process $^4He(e,e'p)X$ have been
investigated. Calculations have been performed including all nuclear responses
in (1) and using the free nucleon structure function of ref. \cite{GHR82};
the relative and c.m. momentum distributions have been taken from ref.
\cite{MOR91}. Before presenting the results, let us remind that at $x<1$ both
backward ($z_2 > 1$) and forward ($z_2 < 1$) nucleon emissions are possible,
whereas at $x>1$ the spectator nucleon is mainly emitted in the forward
hemisphere ($z_2 < 1$); as a matter of fact, at $x>1$ backward nucleon
production is kinematically forbidden within the 2NC model (i.e.,
$\vec{k}_{c.m.} = 0$) and is strongly suppressed within the extended 2NC one
(i.e., $\vec{k}_{c.m.} \neq 0$), because only soft components of
$n^{c.m.}_{N_1N_2}$ contribute to the semi-inclusive nuclear structure function
given by (9). It should also be pointed out that in order to minimize
nucleon emission from quasi-elastic scattering processes on a correlated NN
pair sufficiently high values of $Q^2$ have to be considered. The results of
calculations at $x<1$ are presented in figs. 1 and 2 in the kinematics  region
$0.3 ~ GeV/c < p_2 < 0.7 ~ GeV/c$, where the use of a non relativistic
description of nuclear structure is well grounded. The behaviour of the cross
section as a function of the kinetic energy $T_2$ of the detected nucleon, is
governed by the relative and c.m. momentum distributions of the correlated pair
(particularly at low values of $x$) and by the nucleon structure function
(particularly at high values of $x$). From fig. 1 it can be seen that forward
emission is not sensitive to the two-nucleon spectral function, whereas such a
sensitivity is present for backward emission. This is due to the fact that the
condition $z_1 \simeq 2 - z_2 - (k_{c.m.})_{\|} / M\geq x$ can be satisfied for
forward nucleon emission ($z_2 < 1$) at any values of $p_2$ both with $k_{c.m.}
= 0$ and $k_{c.m.} \neq 0$, whereas for backward nucleon emission ($z_2 > 1$)
and sufficiently large values of $p_2$ it requires that $k_{c.m.} \neq 0$.
	In order to obtain reliable information on the structure function of a bound
nucleon the following ratio has been defined (cf. ref. \cite{CS93})
 \be
    R^{s.i.}(x, x'; \vec{p}_2) = {F^A_2(x, \vec{p}_2) \over F^A_2(x',
    \vec{p}_2)} \cdot {\bar{F}^N_2(x') \over \bar{F}^N_2(x)}
 \ee
with $\bar{F}^N_2 = (F^p_2 + F^n_2) / 2$ and $F^A_2(x, \vec{p}_2) = F^{A,3q}_2
(x, \vec{p}_2)$ given by (9). It turns out that the ratio $R^{s.i.}$ is
almost completely independent on the particular behaviour of the relative and
c.m. momentum distributions and, therefore, it is governed by the behaviour of
the structure function of a bound nucleon.  The ratio, calculated using the
free nucleon structure function in (9), is shown in fig. 2 for backward and
forward nucleon emission and $x \geq x' = 0.3$.  The $x$ dependence of the
calculated ratio sharply reflects the behaviour of the rescaling of the nucleon
structure function: as a matter of fact, for forward nucleon emission ($z_2 <
1$), $x / z_1$ is always less than $x$ and decreases with increasing $p_2$, so
that the ratio is almost constant for small values of $x$ and increases with
increasing $p_2$;  vice versa, for backward nucleon emission ($z_2 > 1$), $x /
z_1$ is  always larger than $x$ and increases with increasing $p_2$, so that
the ratio is less than one and decreases with increasing $p_2$. From fig. 2 it
can be seen that the ratio $R^{s.i.}$ exhibits an appreciable sensitivity upon
the value of $p_2$; therefore, it is clear that any medium-dependent momentum
effect on quark distributions in nuclei can be investigated at forward and
backward nucleon emissions. To sum up, the measurement of the semi-inclusive
cross section at $x<1$ can represent a valuable tool to detect any binding and
momentum dependencies of the nucleon structure function in the medium, and,
consequently, such a measurement can provide non trivial information on the
origin of the EMC effect (cf. ref. \cite{DOU91}). The results presented show
that quantitative predictions for backward nucleon emissions require a careful
treatment of nuclear effects.

\indent At $x > 1 + k_F / M \sim 1.3$ (where $k_F$ is the Fermi momentum) the
semi-inclusive DIS cross section strongly depends upon nuclear effects. This is
illustrated in fig. 3, where the results obtained with the Spectral Functions
given by (4) and (5), are compared at $x=1.2$ and $x=1.5$. It can be
seen that  with increasing $T_2$ the cross section drops out by orders of
magnitude; this is a typical nuclear effect: as a matter of fact, for $T_2 >
250 ~ MeV$ ($p_2 > 0.7 ~ GeV/c$) the cross section mainly follows the behaviour
of the light-cone momentum distributions in the medium rather than the
behaviour of the nucleon structure function; it appears therefore that the
cross section at $x>1$ and $0.3 ~ GeV/c < p_2 < 0.7 ~ GeV/c$ can be used to
investigate quark distributions in nuclei, whereas at higher values of $p_2$ it
provides information on the  short range structure of the two-nucleon system
interacting with the incoming lepton. It should be pointed out that the results
presented have been obtained using the free nucleon structure function;
therefore, any deviation from the behaviour shown in figs. 1-3, could be
ascribed to medium effects on the quark distributions of a bound nucleon.

\indent The predictions of the 2NC and extended 2NC models are compared in fig.
4 with the analysis \cite{MAT89} of the existing experimental data on single
proton backward production in (anti)neutrino charged current interaction with
neon target (Big European Bubble Chamber WA59 collaboration). It can be seen
that, according to (7) the 2NC model predicts for the mean value $<z_1>$ of
the light-cone momentum of the struck nucleon a value $\sim 2 - z_2$
\cite{FS81}, whereas the extended 2NC model predicts a mean value $<z_1>$
larger than $2 - z_2$, in better agreement with the experimental data. The
larger value of $<z_1>$ can be explained as follows: within the 2NC model
(i.e., the correlated NN pair at rest) only $|\vec{k}_1| = |\vec{k}_2|$ is
possible, whereas within the extended 2NC model one can have both $|\vec{k}_1|
< |\vec{k}_2|$ and  $|\vec{k}_1| > |\vec{k}_2|$. The former case is favored,
because it corresponds to smaller values of the initial relative momentum of
the correlated pair and  of the argument of the nucleon structure function.
{}From fig. 4 it can be seen that the difference between the values $<z_1>$
predicted by the two models is not larger than $\sim 0.25$; such a difference
corresponds to values of $k_{c.m.} \sim \sqrt{<k^2_{c.m.}>_{soft}} \sim k_F
\sim 0.25 ~ GeV/c$. It should be pointed  out that in single proton backward
production nuclear effects arising from the Fermi motion of the c.m. of the
correlated NN pair would reduce the role of competitive mechanisms, like, e.g.,
intranuclear cascade and few-nucleon correlations advocated in refs.
\cite{MAT89} and \cite{FS81}, respectively.

\indent The results presented could in principle be modified by mechanisms
different from the spectator one or by the breakdown of the impulse
approximation. In ref. \cite{CS93} the sensitivity of backward and forward
nucleon emissions to the effects of the so-called target fragmentation of the
struck nucleon and the final state interaction of the recoiling nucleon with
the residual (A-2)-nucleon system, has been analyzed; it has been found that
for $0.3 ~ GeV/c < p_2 < 0.7 ~ GeV/c$ (i.e., in the kinematics region of
interest for the spectator mechanism) ~ i) the effects from target
fragmentation of the struck nucleon play a negligible role and do not affect at
all backward nucleon production, and ~ ii) the contributions from the
rescattering of the recoiling nucleon can modify the magnitude of the cross
section without changing in a significant way its dependence upon the kinetic
energy of the detected nucleon.

\vspace{0.5cm}

{\bf 3.	The cross section for the process A(e,e'N)X: six-quark clusters}

\vspace{0.5cm}

\indent Let us now consider the possibility that the nucleons of a correlated
NN pair can lose their identity at short separations; in particular, let us
consider the extreme case in which the short-range structure of the hadronic
cluster interacting with the incoming electron is entirely given by a $6q$ bag
(i.e., $P^{(6q)} = 1$). Within a convolution approach the semi-inclusive
nuclear structure function can be written as
 \be
    F^{A,6q}_2(x, \vec{p}_2) = {A \over 2} S_1 \sum_{\beta} \int_{{x
    \over 2}}^{{M_A \over 2M}} dz_{c.m.} ~ z_{c.m.} ~ f_{\beta}(z_{c.m.}) ~
    \tilde{F}^{\beta (N_2)}_2({x \over 2z_{c.m.}}, {z_2 \over 2z_{c.m.} - 2},
    {\vec{p}^{\perp}_2})
 \ee
where $\beta = (u^2d^4, u^3d^3, u^4d^2) = ([nn],[np],[pp])$ identifies the type
of 6q cluster, $f_{\beta} (z_{c.m.})$ is the light-cone momentum distribution
describing the c.m. motion  of the 6q cluster in the medium, $\tilde{F}^{\beta
(N_2)}_2 (\xi, \varsigma, \vec{p}^{\perp}_2)$ is the fragmentation structure
function of the struck 6q cluster producing a nucleon $N_2$,
$\vec{p}^{\perp}_2$
is the transverse component of the momentum $\vec{p}_2$ of the detected nucleon
with respect to $\vec{q}$ (note that in (11) the mass of the $6q$ cluster
has been assumed to  be 2M). Following ref. \cite{CAR91} it is assumed that
$\tilde{F}^{\beta (N_2)}_2 (\xi, \varsigma, \vec{p}^{\perp}_2)$ factorises
into the structure function of the $6q$ bag, $F^{\beta}_2(\xi)$, and a
fragmentation function $D^{(N_2)}_{(5q)} (\varsigma, \vec{p}^{\perp}_2)$
describing the hadronization of the resulting intermediate $5q$ state.
However, since $D^{(N_2)}_{(5q)} (\varsigma, \vec{p}^{\perp}_2)$ is poorly
known, a simplified form suggested by quark counting rules \cite{QCR} will be
adopted; thus, the final expression of the fragmentation structure function
$\tilde{F}^{\beta (N_2)}_2 (\xi, \varsigma, \vec{p}^{\perp}_2)$ reads as
follows
 \be
    \tilde{F}^{\beta (N_2)}_2 (\xi, \varsigma, \vec{p}^{\perp}_2) = F^{\beta}_2
    (\xi) ~ {\rho_{\perp}(\vec{p}^{\perp}_2) \over E_2} ~ \gamma_{(5q)}
    ~ \varsigma (1 - \varsigma)^3 ~ \Theta(1- \varsigma)
 \ee
where $\rho_{\perp}$ is the transverse momentum distribution of the fragments.
In (12) $\gamma_{(5q)}$ is a constant not known from quark counting rules,
whose value governs the probability (given by $\gamma_{(5q)} \int_0^1
d\varsigma (1 - \varsigma)^3$) of the break-up of the $5q$ system into mesons
or baryons. In the calculations a $50 \%$ probability for the break-up (i.e.,
$\gamma_{(5q)} \int_0^1 d\varsigma (1 - \varsigma)^3 = 0.5$), which implies
$\gamma_{(5q)} = 2$, has been assumed on the analogy of the corresponding
break-up probability of the diquark responsible of the target fragmentation of
the struck nucleon (cf. ref. \cite{BFM82}). It should be pointed out that the
value of $\gamma_{(5q)}$ affects only the absolute value of the cross section.
The $6q$ structure function $F^{\beta}_2(\xi)$ contains both valence and sea
quark distributions, for which the parametrizations of ref. \cite{CAR83} have
been adopted. Scaling violation at finite values of $Q^2$ has been accounted
for by using the Natchmann variable \cite{NAT73} instead of the Bjorken one, as
suggested in ref. \cite{YVHP90}. Finally, for the transverse momentum
distribution $\rho_{\perp}$ appearing in (12) the parametrization of ref.
\cite{DER81} has been used.

\indent The relevant nuclear quantity in (11) is the light-cone c.m.
momentum  distribution $f_{\beta}(z_{c.m.})$ of the $6q$ cluster. For its
evaluation the 2NC and the extended 2NC models have been considered; in the
first model the 6q cluster is considered to be at rest with respect to the
(A-2)-nucleon system, leading to $f_{\beta}(z_{c.m.}) = \delta(z_{c.m.} - 1)$;
in the second one the  c.m. momentum distribution of a $6q$ cluster is assumed
to be the same as the one of a correlated NN pair; this means that a $6q$
cluster and a correlated NN pair differ only in their short-range intrinsic
structure. Thus, the extended 2NC model yields
 \be
    f_{\beta=[N_1N_2]}(z_{c.m.}) = \int d\vec{k}_{c.m.} ~ n^{c.m.}_{N_1N_2}
    (k_{c.m.}) ~ \delta(z_{c.m.} - k^+_{c.m.} / 2M)
 \ee
The semi-inclusive nuclear structure function $F^{A,6q}_2(x, \vec{p}_2)$
(see (11)) becomes
 \be
    F^{A,6q}_2(x, \vec{p}_2) = {A \over 2} S_1 {z_2 \over E_2} \rho_{\perp}
    (\vec{p}^{\perp}_2) \gamma_{(5q)} (1 - {z_2 \over 2 - x})^3 \Theta(2 - x -
    z_2) {1 \over 2 - x} \sum_{\beta} F^{\beta}_2({x \over 2})
 \ee
within the 2NC model (i.e., the $6q$ cluster at rest), whereas one has
 \be
    F^{A,6q}_2(x, \vec{p}_2) = {A \over 2} S_1 {z_2 \over E_2} \rho_{\perp}
    (\vec{p}^{\perp}_2) \gamma_{(5q)} \int_{{x + z_2 \over 2}}^{{M_A \over
    2M}} dz_{c.m.} ~ (1 - {z_2 \over 2z_{c.m.} - x})^3 ~ {z_{c.m.} \over
    2z_{c.m.} - x}
    \\ \nonumber
    \sum_{\beta=[N_1N_2]} F^{\beta}_2({x \over 2z_{c.m.}}) \int d\vec{k}_{c.m.}
    ~ n^{c.m.}_{\beta}(|\vec{k}_{c.m.}|) ~ \delta(z_{c.m.} - k^+_{c.m.} / 2M)
 \ee
within the extended 2NC model.

\indent The results of calculations for forward and backward proton emissions
in
case of the process $^4He(e,e'p)X$ are presented in fig. 5 at $x<1$ and in fig.
6 at $x>1$. It can be seen that, as in the case of the spectator mechanism (see
figs. 1 and 3), backward emission at $x<1$ and forward emission at $x>1$ are
largely affected by the c.m. motion of the hadronic cluster, whereas forward
nucleon emission at $x<1$ is insensitive to nuclear effects. It should be
pointed out that, forward nucleon emission both at $x<1$ and $x>1$ is mainly
governed by the short-range structure of the hadronic cluster interacting with
the incoming lepton; as a matter of fact, the differences in the energy
distribution of emitted nucleons due to $3q$ or $6q$ clusters can be ascribed
to the different short-range structure of the multiquark cluster absorbing the
virtual photon (as it is clearly visible by comparing the results of figs. 5(a)
and 6 with those reported in figs. 1(a) and 3, respectively). On the contrary,
backward nucleon production at $x<1$ is dominated by the c.m. momentum
distribution of the hadronic cluster interacting with the incoming electron, so
that in this case the $T_2$ dependence of the cross section is expected to be
quite similar when the virtual photon is absorbed by 3q or 6q clusters, as it
can be seen by comparing the results of fig. 5(b) with those of fig. 1(b). The
effects of virtual photon absorption on $3q$ and $6q$ clusters upon forward and
backward proton emissions are compared in figs. 7 and 8 (note that the
contributions from the target fragmentation of the struck nucleon, estimated as
in ref. \cite{CS93}, are also reported when the virtual photon absorption
occurs
on a nucleon of a correlated NN pair). It can clearly be seen that the effects
from $6q$ clusters are not expected to sharply modify the $T_2$ dependence of
the cross section when $50 ~ MeV < T_2 < 250 ~ MeV$ (i.e., in the kinematics
region of interest for the spectator mechanism); on the contrary, at higher
values of $T_2$ the presence of $6q$ bags may strongly affect the cross section
both at $x<1$ and $x>1$.

\vspace{0.5cm}

{\bf 4.  Summary and Conclusions}

\vspace{0.5cm}

\indent Forward and backward nucleon emissions in semi-inclusive DIS process
$A(\ell, \ell' N)X$ have been investigated assuming that virtual boson
absorption occurs on a hadronic cluster which can be either a two-nucleon
correlated pair or a six-quark bag. Nuclear effects have been taken care of by
adopting the extended two-nucleon correlation model \cite{CS93}, \cite{CSFS91},
which takes into account the binding and the motion of the center of mass of
the hadronic cluster interacting with the incoming lepton. As for the effects
of virtual photon absorption on a nucleon of a correlated NN pair, the main
results of  the analysis can be listed as follows: ~ i) backward emission at
$x<1$ and forward emission at $x>1$ are sensitive to nuclear effects, whereas
forward emission at $x<1$ is not (cf. fig. 1 and fig. 3); as a matter of fact,
a large part of the discrepancy between the predictions of ref. \cite{FS81}
and the experimental data on single proton backward production observed in
neutrino and antineutrino charged current interactions with neon nuclei
\cite{MAT89}, can be ascribed to nuclear effects arising from the c.m. motion
of the correlated NN pair (cf. fig. 4), rather than to intranuclear cascade or
few-nucleon correlation effects; ~ ii) at $x<1$ the momentum and energy
dependencies of quark distributions in nuclei can be investigated (cf. fig. 2),
providing relevant information on the EMC effect; ~ iii) at $x>1$ the cross
section is mainly governed by the high momentum components of nucleon
light-cone momentum distributions, i.e., by the short-range structure of the
correlated NN pair (cf. fig. 3). As for the effects from six-quark clusters,
the main conclusions are as follows: ~ i) nuclear effects are not important in
forward emission at $x<1$ (cf. fig. 5(a)) but become relevant in backward
emission at $x<1$ (cf. fig. 5(b)) and forward emission at $x>1$ (cf. fig. 6); ~
ii) forward emission both at $x<1$ and $x>1$ appears to be the appropriate
kinematics condition for studying multiquark configurations in nuclei provided
$p_2 > 1 ~ (GeV/c)$ (cf. figs. 7 and 8).

\indent In closing, it should be pointed out that our calculations were
performed within the impulse approximation based on the assumption that the
debris produced by the fragmentation of the hit nucleon do not interact with
the nuclear medium. The problem of the final state interaction (FSI)
of the fragments in inclusive scattering at $x<1$ has been recently addressed
by
various groups \cite{KPW94} - \cite{FSI}. Estimates of FSI effects in
semi-inclusive processes have also been recently obtained \cite{NIK92}, which
suggest that for light nuclei, which is the case considered in the present
paper, FSI should play a minor role thanks to the finite formation time of the
dressed hadrons. Moreover, it should be stressed that backward nucleon emission
is not expected to be affected by forward-produced hadrons \cite{DIE94}.

\newpage

{\bf Figure Captions}

\vspace{0.5cm}

Fig. 1. The semi-inclusive DIS cross section for the process $^4He(e,e'p)X$
within the spectator mechanism, in which the virtual photon is absorbed by a
quark belonging to a nucleon of a correlated NN pair and the recoiling nucleon
is emitted and detected in coincidence with the scattered electron. $T_2$ is
the
kinetic energy of the detected proton, emitted forward ($\theta_2 = 25^o$ (a))
and backward ($\theta_2 = 140^o$ (b)), $x$ is the Bjorken scaling variable. The
curves with and without full dots represent the results obtained within the 2NC
(cf. (7)) and the extended 2NC (cf. (9)) models, respectively. Note that for
$T_2 > 50 ~ MeV$ backward proton emissions at $x=0.6$ and $x=0.8$ are
kinematically forbidden within the 2NC model. Calculations have been performed
assuming the following values for the incident electron energy and scattering
angle: $E_e = 20 ~ GeV$ and $\theta_{e'} = 15^o$. The values of the
four-momentum transfer $Q^2$ are $6, 10, 12, 14 ~ (GeV/c)^2$ for $x=0.2, 0.4,
0.6, 0.8$, respectively.

\vspace{0.5cm}

Fig. 2. The ratio of the semi-inclusive nuclear structure function $R^{s.i.}(x,
x'=0.3; \vec{p}_2)$ (see (11)) for the process $^4He(e,e'p)X$  versus the
Bjorken variable $x$, within the spectator mechanism. Protons are emitted
forward at $\theta_2 = 25^o$ (a) and backward at $\theta_2 = 140^o$ (b). The
solid, dashed, dotted and dot-dashed curves correspond to $p_2= 0.3, 0.4,  0.5$
and $0.6 ~ GeV/c$, respectively. Calculations have been performed assuming $E_e
= 30 ~ GeV$ and $Q^2 = 10 ~ (GeV/c)^2$.

\vspace{0.5cm}

Fig. 3. The same as in fig. 1, but for protons emitted forward at $\theta_2 =
25^o$ and $x>1$.

\vspace{0.5cm}

Fig. 4. Mean value $<z_1>$ of the light-cone momentum of the struck nucleon
versus the light-cone momentum $z_2$ of the detected proton. Open and full dots
represent the experimental data \cite{MAT89} on single proton backward
production from antineutrino and neutrino charged current interaction with neon
target, respectively. Dashed and solid lines are the predictions of the 2NC
(cf. (4)) and extended 2NC (cf. (5)) models, respectively.

\vspace{0.5cm}

Fig. 5. The same as in fig. 1, but assuming that the virtual photon is absorbed
by a quark belonging to a $6q$ bag. The curves with and without full dots
represent the results obtained within the 2NC (cf. (14)) and the extended 2NC
(cf. (15)) models, respectively.

\vspace{0.5cm}

Fig. 6. The same as in fig. 5, but for protons emitted forward at $\theta_2 =
25^o$ and $x>1$.  The kinematics conditions are the same as in fig. 3.

\vspace{0.5cm}

Fig. 7. The semi-inclusive DIS cross section for the process $^4He(e,e'p)X$
versus the kinetic energy $T_2$ of the detected proton, emitted forward at
$\theta_2 = 50^o$ (a) and backward at $\theta_2 = 120^o$ (b), when $x=0.6$. The
solid line is the contribution due to virtual photon absorption on a nucleon of
a correlated NN pair (cf. (9)), whereas the solid line with full dots includes
also the effects from the target fragmentation of the struck nucleon, estimated
as in ref. \cite{CS93}. The dashed line is the contribution resulting from
virtual photon absorption on a quark belonging to a $6q$ bag  (cf. (15)). The
kinematics conditions are the same as in figs. 1 and 5.

\vspace{0.5cm}

Fig. 8. The same as in fig. 7, but for protons emitted forward at $\theta_2 =
25^o$ and $x=1.5$. The kinematics conditions are the same as in figs. 3 and 6.

\end{document}